\documentclass[A4,9ppt]{optica-article}

\journal{opticajournal} 

\articletype{Research Article}

\newcommand{\Ut}[1]{\ensuremath{\widetilde{U_{#1}}}}
\newcommand{\Vt}[1]{\ensuremath{\widetilde{V_{#1}}}}
\newcommand{\Rt}{\ensuremath{\widetilde{R}}}

\usepackage{geometry}
\begin{document}

\title{Arbitrary laser frequency modulation algorithm based on iterative on-the-fly deconvolution}

\author{Thierry Chanelière \authormark{1,*}} 
\address{\authormark{1}Univ. Grenoble Alpes, CNRS, Grenoble INP, Institut N\'eel, 38000 Grenoble, France}
\email{\authormark{*}thierry.chaneliere@neel.cnrs.fr}

 

\begin{abstract*}
I present a general laser modulation control algorithm. I implement the LIDAR Frequency Modulated Continuous Wave (FMCW) scheme as a special case of study. My proposal applies to any arbitrary modulation pattern and is based on an iterative algorithm that infers the laser transfer function in order to perform on-the-fly deconvolution. I present an experimental proof-of-principle using an external-cavity diode laser, the accuracy of which I analyse by comparing the obtained frequency response with a targeted modulation pattern. In addition to the FMCW scheme, I am also testing square wave modulations, which are more demanding in terms of bandwidth.
\end{abstract*}

\tableofcontents
\newpage
\section{Introduction}\label{intro}
Direct control of the laser frequency was only really made possible with the advent of semiconductor lasers. Frequency modulation then appeared as a complement to traditional amplitude modulation in new telecom transmission protocols \cite{Zhu2018}. In addition to the modulation bandwidth required for communication applications, the precision of the modulation scheme was quickly considered for signal processing applications. Some of these applications, which have recently emerged in the field of quantum technologies, require strict yet agile control of atomic emitters \cite{babbitt2020microwave, louchet2020telecom}. But it is undoubtedly in the field of LIDARs that research efforts have been the most sustained for distance and velocity measurements involving both large footprint, very high-precision systems \cite{cariou2006laser} and small, highly-integrated modules \cite{Behroozpour2017}. The triangular modulation scheme in the Frequency Modulated Continuous Wave (FMCW) architecture, for which the linearity of the chirp directly dictates the accuracy of the measurements, sums up the difficulty of the exercise between two contradictory injunctions of speed and accuracy: to rapidly cover a large frequency excursion while maintaining good precision \cite{Ayhan2016}.
The experimental effort is at every level. On the hardware itself, to introduce and shape the modulation elements directly into the cavity. Manufacturing capabilities are currently benefiting greatly from the maturity of photonics-on-a-chip platforms \cite{Bowers, Li2022}. The limitations of the integrated optical setup can be partially compensated for by the addition of internal electronics, sometimes also directly on the chip \cite[and references therein]{Martin2018, Hashemi2022}. The aim of these corrections is to shape the laser response so that it faithfully follows the input command, in other words, to design a flat frequency response, sometimes called frequency equalization \cite{alexander1989passive}. This corrective approach by dedicated electronics is not limited to hardware. The transfer function can be characterised and then taken into account in the input modulation function to obtain the desired output modulation scheme \cite{karlsson1999linearization}.
Another approach, which complements the previous one, is less concerned with knowing the transfer function, but applies input corrections (predistortion) iteratively to converge on the desired output pattern \cite{Zhang2019, Cao2021}.
For the most demanding applications in terms of precision, the toolbox of corrections can be completed by a real-time servo-control on a target function \cite{iiyama1996linearizing, crozatier2006phase}.
All these solutions are generally combined altogether to obtain the best performance \cite{Llauze2024}.

Following this qualitative introduction of the different methods, it is interesting to conduct a brief review of the most recent performance. I will focus on the control of the chirp linearity for FMCW based on a few references. The comparison is not always straightforward, partly because the criteria can vary from one article to another and partly because the excursions and repetition rates differ. However, this allows to extract significant orders of magnitude and position my proof of concept in the LIDAR domain.

The two most compelling demonstrations are those by Zhang {\it et al.}\cite{Zhang2019} and Cao {\it et al.}\cite{Cao2021}, with precisions of around 5$\times10^{-5}$ when comparing the root-mean-square (RMS) value of the chirp non-linearity (2.1\,MHz and 1.5\,MHz respectively) to the excursion (49\,GHz and 26\,GHz respectively), with close repetition rates (4\,kHz and 1\,kHz respectively).
Similar results using a comparable method were obtained by Li {\it et al.}\cite{rs14143455} with a precision of 2$\times10^{-4}$ (5\,MHz RMS for an excursion of 30\,GHz at 8.3\, kHz repetition). The type of laser and modulation scheme must also be taken into account and in the end influence the performance, with, for example, a precision of 4$\times10^{-3}$ (0.66\,MHz RMS, excursion of 162.6\,MHz and 10\,kHz repetition) for direct modulation of the injection current \cite{Yokota:22}. The recent industrial efforts are aimed at maintaining sufficient performance, with a precision of 4$\times10^{-4}$ (3\,MHz RMS, excursion of 7.5\,GHz and 10\,kHz repetition), but with a high level of integration for field applications \cite{10.1117/12.3000397}. Although this analysis is not exhaustive, we can see that an accuracy of $10^{-3}$ is beginning to be of interest. This is a good order of magnitude to remember.

My contribution proposes a new iterative predistortion scheme by linking two previously introduced approaches. On the one hand, a step-by-step measurement of the laser's frequency response allows to produce an arbitrary frequency modulation by deconvolution after this static characterisation. On the other hand, using a trial-and-error approach on the predistorted input signal, one can converge on-the-fly toward the desired modulation pattern without really considering the transfer function. Here I'm gaining knowledge of the transfer function that can be inferred from each trial of a given input signal to propose a compensation for the next iteration round. I will therefore refer to this as on-the-fly deconvolution to mix the terminologies associated with the two approaches, static characterisation and trial-and-error iteration.
By comparison with static measurement of the response function, generally on a laboratory test bench, my method can be used to re-estimate occasionally the transfer function when experimental conditions change in the field. It is compatible with the constraints of an on-board system.

It is also noteworthy that I am not making any assumptions about the nature of the transfer function. It may include derivative or integral terms in addition to the proportional one. This is an important point, because the instantaneous frequency of the laser at a given time $t$ depends on the input signal at previous times, which my method can correct for. Conversely, other correction methods that act at time $t$ to correct at the same time $t$ will be inherently limited.

In addition, I have chosen to modulate the diode injection current directly, which avoids the use of an external modulator. This scheme can be adapted to several types of laser.

From a purely signal processing perspective, it is clearly in the medical field that the advantages and limitations of deconvolution approaches have been most extensively studied. It is impossible to cite all of these works, which span several decades and are still being strongly stimulated by the growth in computing power. I will just give a few entry points to a rich bibliography \cite{ostergaard1996high, jerosch2002myocardial, wu2003tracer, fieselmann2011deconvolution}.

The paper is organised as follows. I will explain the principle of the algorithm in the first part (section \ref{sec:methods}). I then present two variants, which consist in finally estimating the transfer function using Newton's method or the secant method in frequency space (section \ref{sec:algo}). To test iterative convergence, I will use two reference modulation patterns: the triangular modulation typical of FMCW and abrupt jumps between two frequencies (square wave modulation). The second, which I call frequency shift (FS) in reference to the frequency-shift keying (FSK) telecom protocol \cite{Zhu2018}, is more demanding in terms of bandwidth.
I will then show that my methods converge experimentally towards the chosen target functions (section \ref{sec:converge}) and I will analyse the obtained frequency accuracy for the different target patterns (sections \ref{sec:analysis} and \ref{sec:compar}). 

\section{Methods}\label{sec:methods}

\subsection{Iterative deconvolution algorithms}\label{sec:algo}
My approach is based on estimating the laser transfer function. Generally speaking, this latter relates the modulation input, called $U(t)$ as a function of time $t$, often expressed in volts, to the laser output frequency shift, $V(t)$, expressed in MHz in our case. The response function, called $R(t)$, can include both the laser's internal dynamics and the associated electronics on the chip or in the current controller. It characterises the system as a black-box. For a linear system, input and output are related by convolution
$\displaystyle
V = R \circledast U
$
where $\circledast$ is the convolution product. I have omitted the variable $t$ for compactness. From the choice of an arbitrary target modulation function $V_T$, my objective is to find the corresponding command $U_T$ by deconvolution, in other words by estimating the response function $R$. To estimate it, we can choose a first guess function $U_0$, then measure the result $V_0$ on the laser frequency. In Fourier space, marked with a tilde, $\displaystyle \frac{\Vt{0}}{\Ut{0}}$ is an estimate of \Rt \, without considering for the moment the possible division by zero. We can then formally expect $\displaystyle \Vt{T} \frac{\Ut{0}}{\Vt{0}}$ to be the function \Ut{T} we are looking for.

My approach is voluntarily simplistic, because it completely ignores the well-documented limitations of raw deconvolution \cite{jansson1997deconvolution}. They can be summarised as follows. The transfer function may have poles (division by zero) and generally takes the form of a low-pass filter, with the response becoming infinitely weak at high frequencies (it tends towards zero). In these ranges, raw deconvolution simply increases the noise. Properly designed filters usually mitigate this loophole. 

I also assume that the response is perfectly linear whatever the input command. To compensate for this, I propose two iterative methods at first and second order to approach the solution.
Despite the well identified limitations of this simple approach (without filtering), I will show that the algorithm converges in practice.

\subsubsection{First-order Newton's method}\label{sec:Newton}

As I was saying, $\displaystyle \Vt{T} \frac{\Ut{0}}{\Vt{0}}$  is a first estimate of the \Ut{T} search function, so it can be the first iteration \Ut{1} of the following recurrence relation

\begin{equation}
\Ut{n+1}=\Vt{T} \frac{\Ut{n}}{\Vt{n}}
\label{eq:iterative_Fourier}
\end{equation}
There is another way of looking at the iterative formula \eqref{eq:iterative_Fourier}, which can be seen as a root-finding problem of the function $f(\widetilde{U})=\widetilde{V}-\Vt{T}$ where $V$ is the frequency output response to the input modulation $U$.

Eq.\eqref{eq:iterative_Fourier} can be formally rewritten as the Newton's expression \cite{JoannaMPapakonstantinou2013}:
\begin{equation}
\Ut{n+1}=\Ut{n} - (\Vt{n}-\Vt{T}) \frac{\Ut{n}}{\Vt{n}}
\label{eq:iterative_Newton}
\end{equation}
where $\displaystyle f(\Ut{n})=\Vt{n}-\Vt{T}$ is an estimate of the test function $f$ at the iteration step $n$ and  $\displaystyle \frac{\Vt{n}}{\Ut{n}}$ its gradient. This is a first-order method since \Ut{n+1} is inferred from \Ut{n} and \Vt{n} that I will call from now simply Newton's method. This can be generalised to the next order with the secant method.

\subsubsection{Second-order secant method}\label{sec:secant}

The secant method differs in the estimation of the gradient and reads as 

\begin{equation}
\Ut{n}=\Ut{n-1} - (\Vt{n-1}-\Vt{T}) \frac{\Ut{n-1}-\Ut{n-2}}{\Vt{n-1}-\Vt{n-2}}
\label{eq:iterative_secant}
\end{equation}

where $\displaystyle \frac{\Ut{n-1}-\Ut{n-2}}{\Vt{n-1}-\Vt{n-2}} $ is the new estimate of the $f$ gradient. In practice, the algorithm uses the two previous iteration steps, so from the first guess function $U_0$, we calculate $U_1$ using Newton's formula \eqref{eq:iterative_Newton}, and then for the following steps the recurrence \eqref{eq:iterative_secant} applies.

\subsubsection{Implementation with digital experimental waveforms}

While the experiment aims to control the input and output signals in the time domain, I propose an algorithm in frequency space, the two being linked by Fourier transform. Using the FFT (fast Fourier transform) to go from one to the other, however efficient it may be, often poses problems of normalisation and padding. One also needs to pay attention to the units when transforming to give physical meaning to the transfer function. Such precautions are not necessary for my algorithm if we take the same samplings for the series $V_n$ (including $V_T$) on one side and $U_n$  on the other side \footnote{In practice, I use the same equally-spaced temporal sampling for all the $V_n$, $V_T$ and $U_n$}. The ratio $\displaystyle \Vt{T} / \Vt{n}$ in Eq.\eqref{eq:iterative_Newton} ensures that $U_{n+1}$ will, by construction, have the same sampling and normalisation as $U_{n}$. The argument is true for both the Newton's and the secant methods. The Fourier transform marked with a tilde can be directly replaced by the FFT of the digital experimental signals.

\subsection{Target modulation patterns}
As discussed in the introduction \ref{intro}, I'll essentially target two modulation patterns: the triangular modulation (FMCW) and square modulation that I call frequency shift (FS).

They are shown in dotted red in figures  \ref{fig:converge_fmcw}  and \ref{fig:converge_fs} (right columns) to give an idea.
Concerning the repetition rate of these two sequences, I propose a period of $T=100\, \mu$s, typical in LIDAR applications \cite{Hashemi2022, Review_Nanophotonics}.
For the frequency excursion, in this same context, the typical frequency range covers hundreds of MHz for distance and velocity ranging of daily life objects \cite{blackmoreinc}. For my proof-of-principle demonstration,  this is out of reach for my laser, an external-cavity diode laser (ECDL), whose current modulation bandwidth has been optimised for low-noise laser locking, not frequency agility. Compared to a diode alone, an ECDL has inherently lower frequency excursion, directly proportional to its free spectral range. I will therefore limit myself to an excursion of 100\,MHz, ($\pm 50$\,MHz, see Fig. \ref{fig:converge_fmcw}, right column) and 50\,MHz, ($\pm 25$\,MHz, see Figs. \ref{fig:converge_fs}, right column) for the FMCW and FS patterns respectively. The corresponding input modulation signals are a few volts in both cases, limited in practice (see section \ref{sec:setup}) by the output amplitude of my waveform generator ($\pm 5$\,V, see for example the right column of Figs. \ref{fig:converge_fs}, where this maximum value is almost reached).

Both diagrams show abrupt changes, with a frequency jump for FS and a slope break for FMCW. It is interesting to study these breaking points by proposing functions that approach a square (for the FS modualtion) and a triangle (for FMCW) in smoother ways (without singularities). The hyperbolic tangent function \textemdash$\tanh$\textemdash\, tends towards the Heaviside step function when its rise time is zero. We can thus decompose FS square modulation pattern as follows:

\begin{equation}
V_T^\mathrm{FS}(t)=\Delta \times \frac{1}{4} \left[\tanh\left(\frac{ t-T/4}{\tau}\right) - \tanh\left(\frac{ t+T/4}{\tau}\right) -1 \right]\label{eq:VT_FS}
\end{equation}
where the parameter $\tau$ is the rise time and $\Delta=50$\,MHz the frequency excursion for FS. As an illustration, I will vary $\tau$ from $200$ ns to $2.6\,\mu$s for a repetition period of $T=100\, \mu$s in the study of section \ref{sec:compar}.

The triangular pattern, as the integral of a square, can therefore be written for FMCW modulation:

\begin{equation}
V_T^\mathrm{FMCW}(t)=\Delta \times \frac{\tau}{T} \left[ \log\left(\cosh \left(\frac{ t-T/4}{\tau}\right)\right) - \log\left(\cosh \left(\frac{ t+T/4}{\tau}\right) \right) -\frac{t}{\tau} \right]\label{eq:VT_FMCW}
\end{equation}
where we recognise $\log(\cosh)$ as the integral of $\tanh$. $\Delta=100$\,MHz the frequency excursion for FMCW.
For the illustrations in figures  \ref{fig:converge_fmcw}  and \ref{fig:converge_fs} (right columns), the rise time $\tau$ is adjusted to $600$ ns.

After this presentation of the methods, it's time to detail their experimental implementation.

\section{Experiments}

I'll start by describing the experiment, which consists of a laser and a frequency discriminator. The associated electronics are used to modulate and record the signals. I'll then show how the algorithm converges in practice. I'll conclude by setting out the criteria for the the comparative analysis in section \ref{sec:compar}.

\subsection{Optoelectronics setup}\label{sec:setup}

As I've already said, the laser is an ECDL whose frequency excursion is intrinsically limited by design. I apply direct injection current modulation to the diode through a T-bias. The modulation sensitivity is about 30MHz/V through a 510 ohm resistor. The DC component is cut off below approximately 100Hz, which does not allow control of the average frequency. This is a technical rather than a fundamental limitation of the algorithm.
The frequency discriminator is a Michelson-type fibre interferometer. It is used to measure the instantaneous frequency directly when the laser is positioned on the side of the fringe. In my case, it is actively maintained by a slow servo-control whose response time is much longer than the repetition rate. So as not to digress in the main article, I discuss this choice more specifically in the Supplemental document (subsection "Short-arm or long-arm interferometer"). I take this opportunity to give a technical description of the interferometer used in the Supplemental document (section "Frequency discriminators") as well as the detection scheme.

The input modulation functions $U_n$ are generated by programing an arbitrary function generator (Tektronix AFG3021B). As a first guess $U_0$, I choose a simple function that looks like the target output pattern, namely triangular and square inputs for the FMCW and FS target output patterns respectively. For a given iteration $U_n$, the output $V_n$ is recorded, and the next step for $U_{n+1}$ is calculated by using the iterative expressions Eq.\eqref{eq:iterative_Newton} or Eq.\eqref{eq:iterative_secant} for the Newton's method \ref{sec:Newton} or second-order secant method \ref{sec:secant} respectively depending on the case of study.

I first test the convergence of the algorithm for the Newton's method in my two situations of interest, i.e. targeting the output FMCW and FS frequency modulation patterns. I will then examine the second-order secant method, rather in section \ref{sec:analysis}, with a comparative analysis of the algorithm variants.

\subsection{Iterative convergence}\label{sec:converge}
I've introduced analytical functions (Eqs. \ref{eq:VT_FS} and \ref{eq:VT_FMCW}) whose rise time can be smoothed, more or less abruptly, to represent the FCMW or FS patterns respectively.

\begin{figure}[htbp]
\centering
\includegraphics[width=.95\columnwidth]{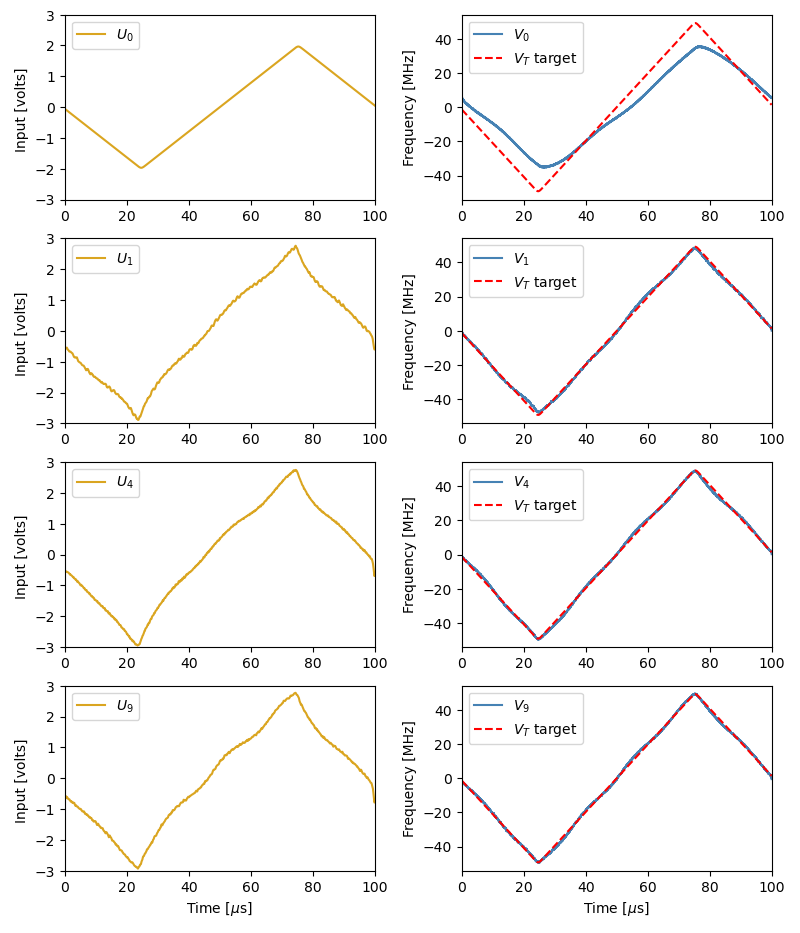} 
\caption{Right column: converge of the algorithm after 10 iterations for the target FMCW pattern $V_T^\mathrm{FMCW}$ with a rise time $\tau=600$\,ns (red dashed line). Left column: I represent the input modulation signal $U_n$ for the first ( $U_0$ as the initial guess, first line), second ($U_1$, second line), fifth ($U_4$, third line ) and tenth ($U_9$, last line) iteration.}
\label{fig:converge_fmcw}
\end{figure}

\begin{figure}[htbp]
\centering
\includegraphics[width=.95\columnwidth]{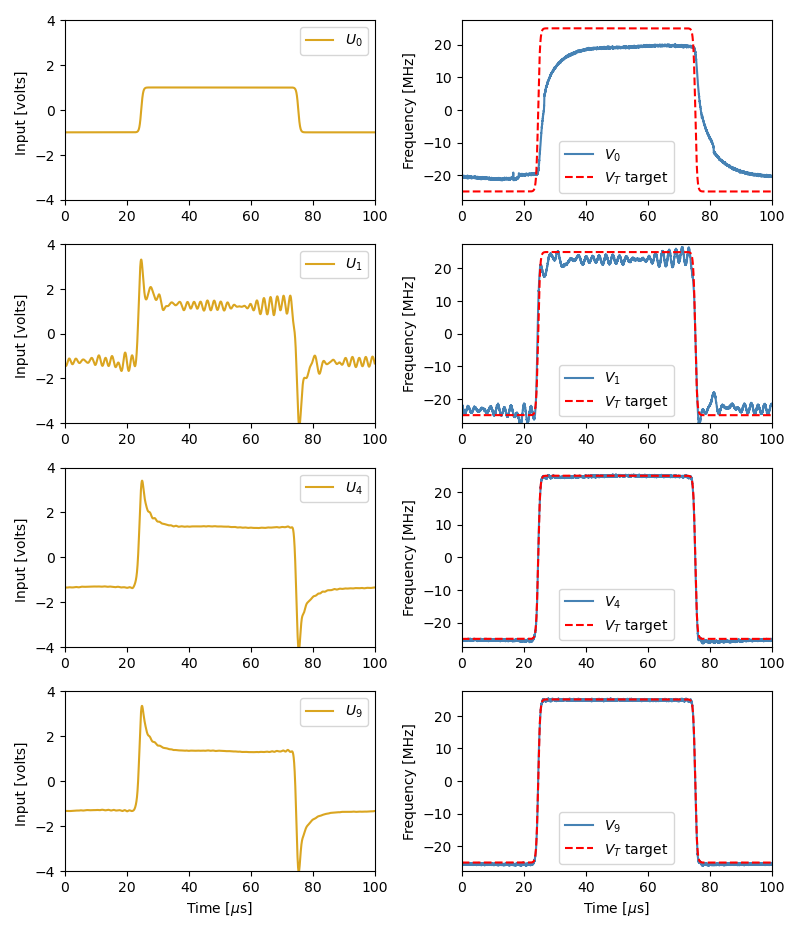} 
\caption{Converge of the algorithm after 10 iterations for the target FS pattern $V_T^\mathrm{FS}$ (red dashed line) with the same representation as Fig. \ref{fig:converge_fmcw} and same rise time $\tau=600$\,ns.}
\label{fig:converge_fs}
\end{figure}

As an illustration, I chose a rise time $\tau=600$\,ns which is already much shorter than the repetition time $T=100\, \mu$s. I will vary $\tau$ later on in section \ref{sec:compar} to investigate the converge in the different conditions. The iterations are represented in Figs. \ref{fig:converge_fmcw} and \ref{fig:converge_fs} for the FCMW or FS patterns respectively using the Newton's method (Eq.\eqref{eq:iterative_Newton}).

Despite the reservations I had when introducing the deconvolution algorithm in \ref{sec:algo}, we can see that it converges without difficulty to the desired pattern.

The presence of possible poles in the response function is not prohibitive to convergence, but it may translate into the presence of well-defined oscillations as can be seen in the two situations FMCW or FS with different consequences. On the one hand, they can persist after iterations, as can be seen in the Fig.\ref{fig:converge_fmcw} with a slight oscillation of period around 20$\mu$s. This is a limitation of the algorithm. On the other hand, they may appear for one iteration, but then disappear simply because the noise at that frequency is gradually reduced by the algorithm, as shown in the Fig.\ref{fig:converge_fs} (2$^\mathrm{nd}$ iteration on $V_1$ with a period of around 2\,$\mu$s, then disappearing).

The visual agreement is satisfactory and convergence is possible without any a priori knowledge of the transfer function and by choosing an initial estimate without any particular care. I will now carry out a more detailed analysis of the iterations by quantifying the evolution of the error between the observed modulation and the target function.

\subsection{Converge analysis}\label{sec:analysis}

To quantify the error with respect to the target function, I choose as the criterion the root mean square deviation (RMSD) between $V_T$ and $V_n$. This can be calculated at each iteration. For information, the acquisition sampling rate is 82\,MHz for a total duration of $T=100\, \mu$s ($2^{13}$ samples per waveform).

In any case, the RMSD value is limited by the frequency noise of the laser. To determine this, we turn off all modulation and measure a RMSD noise floor of 130\,kHz. This is a typical linewidth value for an ECDL with an integration time of $100\, \mu$s \cite{ECDL_noise}, and I will use it as a base reference in the analysis that follows.

\begin{figure}[htbp]
\centering
\includegraphics[width=.95\columnwidth]{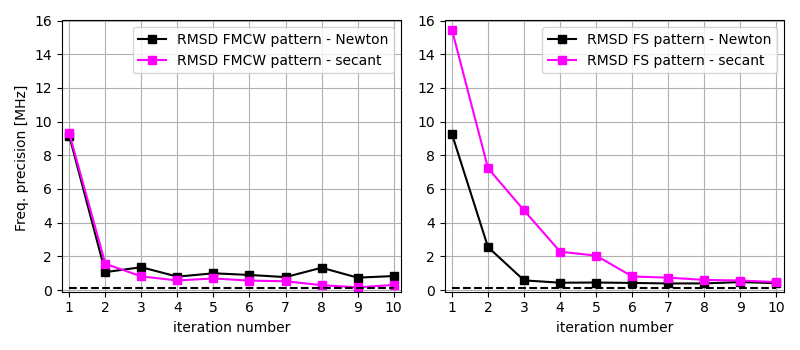} 
\caption{Frequency precision, defined as the RMSD error for the target function, converge after 10 iterations for the FCMW (left panel) and the FS patterns (right panel). For each pattern, I compare the first-order Newton's method (black curve and symbols), already illustrated in Figs. \ref{fig:converge_fmcw} and \ref{fig:converge_fs}, and the second-order secant method (magenta curve and symbols). The 130\,kHz noise floor is represented as a black dashed line. The rise time $\tau=600$\,ns is unchanged. The RMSD values at the last iteration are summarised in Table.\ref{table:rmsd}.}
\label{fig:converge_compar}
\end{figure}

I compare in Fig.\ref{fig:converge_compar} the converge after 10 iterations for the FCMW and the FS patterns. Convergence is assured after 10 iterations, and I have not observed any significant improvement beyond that, regardless of the scheme and method. I take the opportunity to implement and include in this comparison the second-order secant method introduced in \ref{sec:secant}.

All the methods converge correctly towards the desired solution. The secant method is not systematically better, even though it relies on a higher order of approximation. I'll analyse this in more detail later, as it's impossible to generalise from a single example with a common  rise time $\tau=600$\,ns. Nevertheless, we can see that the FMCW target modulation is better controlled by the algorithm than the FS pattern, at least on the first few iterations. This is not surprising, because as I said in the introduction, a square modulation relies on a wider frequency spectrum. The algorithm seems less capable to filter out the noise, which is a possible source of divergence.

For a given target modulation (FMCW or FS) and method (Newton or secant), I use the RMSD value at the 10$^\mathrm{th}$ and last iteration as the convergence criterion \textemdash frequency precision at iteration number 10 in Fig.\ref{fig:converge_compar}\textemdash\,  that I will use in the following. The results are summarised in Table \ref{table:rmsd}.

\begin{table}[htbp]
\caption{Frequency precision of the algorithm at the last iteration (10$^\mathrm{th}$ iteration) in frequency units and normalised to the excursion $\Delta$.}\label{table:rmsd}
 \centering
\begin{tabular}{ |c|c|c| } 
 \hline
RMSD at the 10$^\mathrm{th}$ iteration $\vert$ RMSD/$\Delta$ & FMCW pattern & FS pattern \\ \hline
Newton method & 831\,kHz $\vert$ 8.3$\times10^{-3}$ & 409\,kHz $\vert$ 8.2$\times10^{-3}$\\ \hline
Secant method & 304\,kHz $\vert$ 3.0$\times10^{-3}$ & 475\,kH $\vert$ 9.5$\times10^{-3}$\\ 
 \hline
\end{tabular}
\end{table}

Since the two patterns do not have the same frequency excursion $\Delta$, it is useful to represent the ratio of the frequency error to $\Delta$ as well (see Table \ref{table:rmsd}, where the dimensionless value is RMSD/$\Delta$). We also need to bear in mind that the value of the noise floor is 130\,kHz. We can see that the overall error is a few $10^{-3}$, it remains higher, but comparable to the noise floor.

This is an interesting first step towards applications, particularly in LIDAR, where frequency resolution has a direct impact on distance and speed resolution, a few $10^{-3}$ in this case. Despite the general convergence of the algorithm, there are significant disparities between the four examples, both during the iterations (Fig.\ref{fig:converge_compar}) and in the final value of the error (Table \ref{table:rmsd}). In order to assess noise emergence for a given target function, I propose to vary the spectrum by adjusting the rise time, which acts as a smoothing effect for both the FMCW and FS patterns.

\section{Comparison and discussion}\label{sec:compar}

The  analytical functions (Eqs \ref{eq:VT_FMCW} and \ref{eq:VT_FS}) offer a smoothed version of the FMCW or FS patterns whose rise time $\tau$ can be varied by programming the arbitrary function generator.

\begin{figure}[htbp]
\centering
\includegraphics[width=.95\columnwidth]{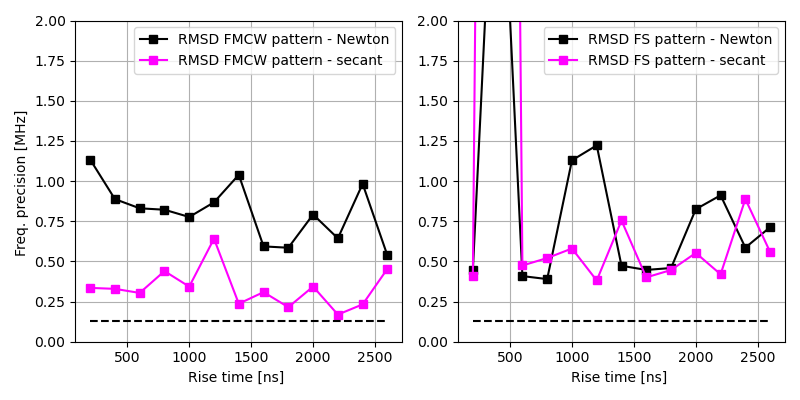} 
\caption{Frequency precision as a function of $\tau$ for each pattern and method with the same colour code as in Fig.\ref{fig:converge_compar}.}
\label{fig:rmsd_vs_rise_time}
\end{figure}

The result is illustrated in Fig.\ref{fig:rmsd_vs_rise_time} for the different patterns and methods.

The FMCW pattern is relatively robust regardless of the value of $\tau$, even though some variability is observed. More generally, I confirm the trend observed for the single example with $\tau=600$ ns. The secant method systematically performs better than Newton's method and approaches the noise floor.

The FS pattern is clearly more chaotic. For $\tau=400$ ns, convergence is not completely assured and there is a significant disparity between the precision values. For the sake of completeness, I will take the time to illustrate this counterexample, which shows the limitations of the algorithm in the Supplemental document (section "A counter-example to convergence").
More generally, we see the same trend as in the table \ref{table:rmsd}. The secant method does not really perform better than Newton's method, this trend does not change if the number of iterations is increased. The relative frequency accuracy is around $10^{-2}$. It is worth noting again that the FS pattern is particularly demanding with abrupt temporal changes. Here, I make a first step significant towards stabilising the modulations without prior knowledge of the response function which is inferred from the measurement and exploited iteratively.


\section{Conclusion and perspectives}\label{sec:ccl}

I propose an on-the-fly deconvolution method to control laser frequency modulation, for which I provide proof-of-principle using an ECDL. I show that the iterative algorithm converges globally, even though certain rare temporal shapes should be avoided as convergence is not guaranteed for them. This is not surprising; one might even expect worse, since the response function, without seeking to plot it here, may have poles in the spectral domain that act as points of attraction where the noise is amplified.

For the FMCW modulation scheme, which is of interest for LIDAR, I obtain an accuracy of few $10^{-3}$, whereas the laser noise floor is $10^{-3}$. My algorithm is therefore a simple and effective first step towards modulation control. The algorithmic complexity is low since the method involves linear combinations of FFTs that can be calculated quickly even in an on-board system. My approach is therefore not necessarily restricted to a laboratory test bench, but can be effectively implemented in operational environment for recalibrating devices.

My experimental demonstration is limited to reduced frequency excursions ($<100$\,MHz) imposed by the laser used, an ECDL whose current modulation range is intrinsically limited due to the extended length of the cavity. A distributed-feedback laser (DFB), for which the method could be used, would have a much larger excursion, more compatible with applications ($\sim 1$\,GHz).

I chose to present the raw method in order to demonstrate its advantages and limitations. One possible improvement would obviously be to apply spectral domain filtering with two objectives. On the one hand, to try to achieve convergence in situations where it is clearly difficult as illustrated the counterexample in the Supplemental document (section "A counter-example to convergence"). Secondly, when convergence is assured, to improve the frequency precision in order to approach the laser noise limit. All kinds of processing filters can be imagined, such as low-pass frequency filters, but it would be more interesting to identify and eliminate the poles. To do this, it is not necessary to fully measure the response function. We could extend our blind approach \textemdash without plotting the the response function\textemdash\, by favouring the largest components in the spectral domain and canceling out the others, in Eq.\ref{eq:iterative_Fourier} for example. This would correspond to a cut-off in the singular value decomposition (SVD) whose arrangement is  actually ordered (largest first). The Fourier transform is actually a special case of SVDs, which serve as the basis for many deconvolution methods.

To conclude on the contextual elements listed in the introduction, the method I propose is not an end in itself, but is intended to be combined with other hardware and software techniques aimed at improving laser modulation controls.

\begin{backmatter}
\bmsection{Funding}
The author acknowledge support from the Plan France 2030 project QMEMO No. ANR-22-PETQ-0010.

\bmsection{Acknowledgment}
I would like to thank Vincent Pureur, Rebecca El Koussaifi and Jean-Pierre Cariou from VAISALA France for the discussions in the context of LIDAR and for their encouragement.

\bmsection{Disclosures}
The author declares no conflicts of interest.

\bmsection{Data availability}
Data underlying the results presented in this paper are not publicly available at this time but may be obtained from the authors upon reasonable request.

\bmsection{Supplemental document}
See Supplemental document for supporting content.

\end{backmatter}

\bibliography{FM_control_algo_bib_resub}{}

@article{Zhu2018,
  title = {Directly Modulated Semiconductor Lasers},
  volume = {24},
  ISSN = {1558-4542},
  url = {http://dx.doi.org/10.1109/JSTQE.2017.2720959},
  DOI = {10.1109/jstqe.2017.2720959},
  number = {1},
  journal = {IEEE Journal of Selected Topics in Quantum Electronics},
  publisher = {Institute of Electrical and Electronics Engineers (IEEE)},
  author = {Zhu,  Ning Hua and Shi,  Zhan and Zhang,  Zhi Ke and Zhang,  Yi Ming and Zou,  Can Wen and Zhao,  Ze Ping and Liu,  Yu and Li,  Wei and Li,  Ming},
  year = {2018},
  month = jan,
  pages = {1–19}
}

@inproceedings{babbitt2020microwave,
  title={Microwave photonic processing with spatial-spectral holographic materials},
  author={Babbitt, Wm Randall},
  booktitle={Optical, Opto-Atomic, and Entanglement-Enhanced Precision Metrology II},
  volume={11296},
  pages={258--268},
  year={2020},
  organization={SPIE}
}

@article{louchet2020telecom,
  title={Telecom wavelength optical processor for wideband spectral analysis of radiofrequency signals},
  author={Louchet-Chauvet, Anne and Berger, Perrine and Nouchi, Pascale and Dolfi, Daniel and Ferrier, Alban and Goldner, Philippe and Morvan, Lo{\"\i}c},
  journal={Laser Physics},
  volume={30},
  number={6},
  pages={066203},
  year={2020},
  publisher={IOP Publishing}
}

@article{Behroozpour2017,
  title = {Lidar System Architectures and Circuits},
  volume = {55},
  ISSN = {0163-6804},
  url = {http://dx.doi.org/10.1109/MCOM.2017.1700030},
  DOI = {10.1109/mcom.2017.1700030},
  number = {10},
  journal = {IEEE Communications Magazine},
  publisher = {Institute of Electrical and Electronics Engineers (IEEE)},
  author = {Behroozpour,  Behnam and Sandborn,  Phillip A. M. and Wu,  Ming C. and Boser,  Bernhard E.},
  year = {2017},
  month = oct,
  pages = {135–142}
}

@article{cariou2006laser,
  title={Laser source requirements for coherent lidars based on fiber technology},
  author={Cariou, Jean-Pierre and Augere, B{\'e}atrice and Valla, Matthieu},
  journal={Comptes Rendus Physique},
  volume={7},
  number={2},
  pages={213--223},
  year={2006},
  publisher={Elsevier}
}

@article{Ayhan2016,
  title = {Impact of Frequency Ramp Nonlinearity,  Phase Noise,  and SNR on FMCW Radar Accuracy},
  volume = {64},
  ISSN = {1557-9670},
  url = {http://dx.doi.org/10.1109/TMTT.2016.2599165},
  DOI = {10.1109/tmtt.2016.2599165},
  number = {10},
  journal = {IEEE Transactions on Microwave Theory and Techniques},
  publisher = {Institute of Electrical and Electronics Engineers (IEEE)},
  author = {Ayhan,  Serdal and Scherr,  Steffen and Bhutani,  Akanksha and Fischbach,  Benjamin and Pauli,  Mario and Zwick,  Thomas},
  year = {2016},
  month = oct,
  pages = {3290–3301}
}

@ARTICLE{Bowers,
  author={Komljenovic, Tin and Davenport, Michael and Hulme, Jared and Liu, Alan Y. and Santis, Christos T. and Spott, Alexander and Srinivasan, Sudharsanan and Stanton, Eric J. and Zhang, Chong and Bowers, John E.},
  journal={Journal of Lightwave Technology}, 
  title={Heterogeneous Silicon Photonic Integrated Circuits}, 
  year={2016},
  volume={34},
  number={1},
  pages={20-35},
  doi={10.1109/JLT.2015.2465382}}

@article{Li2022,
  title = {Integrated Pockels laser},
  volume = {13},
  ISSN = {2041-1723},
  url = {http://dx.doi.org/10.1038/s41467-022-33101-6},
  DOI = {10.1038/s41467-022-33101-6},
  number = {1},
  journal = {Nature Communications},
  publisher = {Springer Science and Business Media LLC},
  author = {Li,  Mingxiao and Chang,  Lin and Wu,  Lue and Staffa,  Jeremy and Ling,  Jingwei and Javid,  Usman A. and Xue,  Shixin and He,  Yang and Lopez-rios,  Raymond and Morin,  Theodore J. and Wang,  Heming and Shen,  Boqiang and Zeng,  Siwei and Zhu,  Lin and Vahala,  Kerry J. and Bowers,  John E. and Lin,  Qiang},
  year = {2022},
  month = sep 
}

@inproceedings{Hashemi2022,
  title = {A Review of Silicon Photonics LiDAR},
  url = {http://dx.doi.org/10.1109/CICC53496.2022.9772845},
  DOI = {10.1109/cicc53496.2022.9772845},
  booktitle = {2022 IEEE Custom Integrated Circuits Conference (CICC)},
  publisher = {IEEE},
  author = {Hashemi,  Hossein},
  year = {2022},
  month = apr,
  pages = {1–8}
}

@article{alexander1989passive,
  title={Passive equalization of semiconductor diode laser frequency modulation},
  author={Alexander, Stephen B and Welford, D and Marquis, DVL},
  journal={Journal of lightwave technology},
  volume={7},
  number={1},
  pages={11--23},
  year={1989},
  publisher={IEEE}
}

@article{Martin2018,
  title = {Photonic Integrated Circuit-Based FMCW Coherent LiDAR},
  volume = {36},
  ISSN = {1558-2213},
  url = {http://dx.doi.org/10.1109/JLT.2018.2840223},
  DOI = {10.1109/jlt.2018.2840223},
  number = {19},
  journal = {Journal of Lightwave Technology},
  publisher = {Institute of Electrical and Electronics Engineers (IEEE)},
  author = {Martin,  Aude and Verheyen,  Peter and De Heyn,  Peter and Absil,  Philippe and Feneyrou,  Patrick and Bourderionnet,  Jerome and Dodane,  Delphin and Leviandier,  Luc and Dolfi,  Daniel and Naughton,  Alan and O’Brien,  Peter and Spuessens,  Thijs and Baets,  Roel and Lepage,  Guy},
  year = {2018},
  month = oct,
  pages = {4640–4645}
}

@article{karlsson1999linearization,
  title={Linearization of the frequency sweep of a frequency-modulated continuous-wave semiconductor laser radar and the resulting ranging performance},
  author={Karlsson, Christer J and Olsson, Fredrik {\AA}A},
  journal={Applied optics},
  volume={38},
  number={15},
  pages={3376--3386},
  year={1999},
  publisher={Optica Publishing Group}
}

@article{iiyama1996linearizing,
  title={Linearizing optical frequency-sweep of a laser diode for FMCW reflectometry},
  author={Iiyama, Koichi and Wang, Lu-Tang and Hayashi, Ken-Ichi},
  journal={Journal of lightwave technology},
  volume={14},
  number={2},
  pages={173--178},
  year={1996},
  publisher={IEEE}
}

@article{crozatier2006phase,
  title={Phase locking of a frequency agile laser},
  author={Crozatier, Vincent and Gorju, Guillaume and Bretenaker, Fabien and Le Gou{\"e}t, Jean-Louis and Lorger{\'e}, Ivan and Gagnol, Claude and Ducloux, Eric},
  journal={Applied physics letters},
  volume={89},
  number={26},
  year={2006},
  publisher={AIP Publishing}
}

@article{Zhang2019,
  title = {Laser frequency sweep linearization by iterative learning pre-distortion for FMCW LiDAR},
  volume = {27},
  ISSN = {1094-4087},
  url = {http://dx.doi.org/10.1364/OE.27.009965},
  DOI = {10.1364/oe.27.009965},
  number = {7},
  journal = {Optics Express},
  publisher = {Optica Publishing Group},
  author = {Zhang,  Xiaosheng and Pouls,  Jazz and Wu,  Ming C.},
  year = {2019},
  month = mar,
  pages = {9965}
}

@article{Cao2021,
  title = {Highly efficient iteration algorithm for a linear frequency-sweep distributed feedback laser in frequency-modulated continuous wave lidar applications},
  volume = {38},
  ISSN = {1520-8540},
  url = {http://dx.doi.org/10.1364/JOSAB.430605},
  DOI = {10.1364/josab.430605},
  number = {10},
  journal = {Journal of the Optical Society of America B},
  publisher = {Optica Publishing Group},
  author = {Cao,  Xianyi and Wu,  Kan and Li,  Chao and Zhang,  Guangjin and Chen,  Jianping},
  year = {2021},
  month = jul,
  pages = {D8}
}

@article{Llauze2024,
  title = {Versatile,  fast,  and accurate frequency excursions with a semiconductor laser},
  volume = {63},
  ISSN = {2155-3165},
  url = {http://dx.doi.org/10.1364/AO.522789},
  DOI = {10.1364/ao.522789},
  number = {19},
  journal = {Applied Optics},
  publisher = {Optica Publishing Group},
  author = {Llauze,  Thomas and Montjovet-Basset,  Félix and Louchet-Chauvet,  Anne},
  year = {2024},
  month = jun,
  pages = {5192}
}

@book{jansson1997deconvolution,
  title={Deconvolution of Images and Spectra},
  author={Jansson, P.A.},
  isbn={9780123802224},
  lccn={96027097},
  year={1997},
  publisher={Academic Press}
}

@article{JoannaMPapakonstantinou2013,
  title = {Origin and Evolution of the Secant Method in One Dimension},
  volume = {120},
  ISSN = {0002-9890},
  url = {http://dx.doi.org/10.4169/amer.math.monthly.120.06.500},
  DOI = {10.4169/amer.math.monthly.120.06.500},
  number = {6},
  journal = {The American Mathematical Monthly},
  publisher = {Informa UK Limited},
  author = {Joanna M. Papakonstantinou and Richard A. Tapia},
  year = {2013},
  pages = {500}
}

@article{Review_Nanophotonics,
author = {Li, Nanxi and Ho, Chong Pei and Xue, Jin and Lim, Leh Woon and Chen, Guanyu and Fu, Yuan Hsing and Lee, Lennon Yao Ting},
title = {A Progress Review on Solid-State LiDAR and Nanophotonics-Based LiDAR Sensors},
journal = {Laser \& Photonics Reviews},
volume = {16},
number = {11},
pages = {2100511},
doi = {https://doi.org/10.1002/lpor.202100511},
url = {https://onlinelibrary.wiley.com/doi/abs/10.1002/lpor.202100511},
eprint = {https://onlinelibrary.wiley.com/doi/pdf/10.1002/lpor.202100511},
year = {2022}
}

@INPROCEEDINGS{blackmoreinc,
  author={Kadlec, Emil A. and Barber, Zeb W. and Rupavatharam, Krishna and Angus, Edward and Galloway, Ryan and Rogers, Evan M. and Thornton, Joshua and Crouch, Stephen},
  booktitle={2019 24th OptoElectronics and Communications Conference (OECC) and 2019 International Conference on Photonics in Switching and Computing (PSC)}, 
  title={Coherent Lidar for Autonomous Vehicle Applications}, 
  year={2019},
  volume={},
  number={},
  pages={1-3},
  doi={10.23919/PS.2019.8817713}}

@ARTICLE{ECDL_noise,
  author={van Exter, M.P. and Kuppens, S.J.M. and Woerdman, J.P.},
  journal={IEEE Journal of Quantum Electronics}, 
  title={Excess phase noise in self-heterodyne detection}, 
  year={1992},
  volume={28},
  number={3},
  pages={580-584},
  doi={10.1109/3.124980}}

@article{ostergaard1996high,
  title={High resolution measurement of cerebral blood flow using intravascular tracer bolus passages. Part I: Mathematical approach and statistical analysis},
  author={{\O}stergaard, Leif and Weisskoff, Robert M and Chesler, David A and Gyldensted, Carsten and Rosen, Bruce R},
  journal={Magnetic resonance in medicine},
  volume={36},
  number={5},
  pages={715--725},
  year={1996},
  publisher={Wiley Online Library}
}

@article{jerosch2002myocardial,
  title={Myocardial blood flow quantification with MRI by model-independent deconvolution},
  author={Jerosch-Herold, Michael and Swingen, Cory and Seethamraju, Ravi Teja},
  journal={Medical physics},
  volume={29},
  number={5},
  pages={886--897},
  year={2002},
  publisher={Wiley Online Library}
}

@article{wu2003tracer,
  title={Tracer arrival timing-insensitive technique for estimating flow in MR perfusion-weighted imaging using singular value decomposition with a block-circulant deconvolution matrix},
  author={Wu, Ona and {\O}stergaard, Leif and Weisskoff, Robert M and Benner, Thomas and Rosen, Bruce R and Sorensen, A Gregory},
  journal={Magnetic Resonance in Medicine: An Official Journal of the International Society for Magnetic Resonance in Medicine},
  volume={50},
  number={1},
  pages={164--174},
  year={2003},
  publisher={Wiley Online Library}
}

@article{fieselmann2011deconvolution,
  title={Deconvolution-based CT and MR brain perfusion measurement: theoretical model revisited and practical implementation details},
  author={Fieselmann, Andreas and Kowarschik, Markus and Ganguly, Arundhuti and Hornegger, Joachim and Fahrig, Rebecca},
  journal={International Journal of Biomedical Imaging},
  volume={2011},
  number={1},
  pages={467563},
  year={2011},
  publisher={Wiley Online Library}
}

@article{Jiang:10,
author = {Haifeng Jiang and Fabien K\'{e}f\'{e}lian and Pierre Lemonde and Andr\'{e} Clairon and Giorgio Santarelli},
journal = {Opt. Express},
number = {4},
pages = {3284--3297},
publisher = {Optica Publishing Group},
title = {An agile laser with ultra-low frequency noise and high sweep linearity},
volume = {18},
month = {Feb},
year = {2010},
url = {https://opg.optica.org/oe/abstract.cfm?URI=oe-18-4-3284},
doi = {10.1364/OE.18.003284},
}

@article{Yokota:22,
author = {Nobuhide Yokota and Hiroki Kiuchi and Hiroshi Yasaka},
journal = {Opt. Express},
number = {7},
pages = {11693--11703},
publisher = {Optica Publishing Group},
title = {Directly modulated optical negative feedback lasers for long-range FMCW LiDAR},
volume = {30},
month = {Mar},
year = {2022},
url = {https://opg.optica.org/oe/abstract.cfm?URI=oe-30-7-11693},
doi = {10.1364/OE.452284},
}

@Article{rs14143455,
AUTHOR = {Li, Peng and Zhang, Yating and Yao, Jianquan},
TITLE = {Rapid Linear Frequency Swept Frequency-Modulated Continuous Wave Laser Source Using Iterative Pre-Distortion Algorithm},
JOURNAL = {Remote Sensing},
VOLUME = {14},
YEAR = {2022},
NUMBER = {14},
ARTICLE-NUMBER = {3455},
URL = {https://www.mdpi.com/2072-4292/14/14/3455},
ISSN = {2072-4292},
DOI = {10.3390/rs14143455}
}

@inproceedings{10.1117/12.3000397,
author = {Vincent Cardin and Daniel Robin and Sylvain Boudreau and Guy Rousseau and Simon Ayotte and Marie-Claude Vallieres Riendeau and Patrick Larochelle and Fran{\c{c}}ois Costin and {\'E}mile Girard-Desch{\^e}nes and Philippe Chr{\'e}tien and Guillaume Brochu and Patrick Dufour and S{\'e}bastien Desch{\^e}nes and Katherine L{\'e}gar{\'e} and Mathieu Faucher and Mohamed Rahim and Grzegorz Pakulski and Muhammad Mohsin and Darren Goodchild and Philip Waldron and Bernard Paquette and Omid Salehzadeh Einabad and Daniel Poitras},
title = {{Narrow-linewidth semiconductor laser with highly-linear frequency modulation response for coherent sensing}},
volume = {12905},
booktitle = {Novel In-Plane Semiconductor Lasers XXIII},
editor = {Alexey A. Belyanin and Peter M. Smowton},
organization = {International Society for Optics and Photonics},
publisher = {SPIE},
pages = {129050F},
year = {2024},
doi = {10.1117/12.3000397},
URL = {https://doi.org/10.1117/12.3000397}
}

\newpage
\appendix
\addcontentsline{toc}{section}{Supplemental document:}
\title{Supplemental document}

\section{A counter-example to convergence}\label{appendix:counterexample}

\begin{figure}[htbp]
\centering
\includegraphics[width=.95\columnwidth]{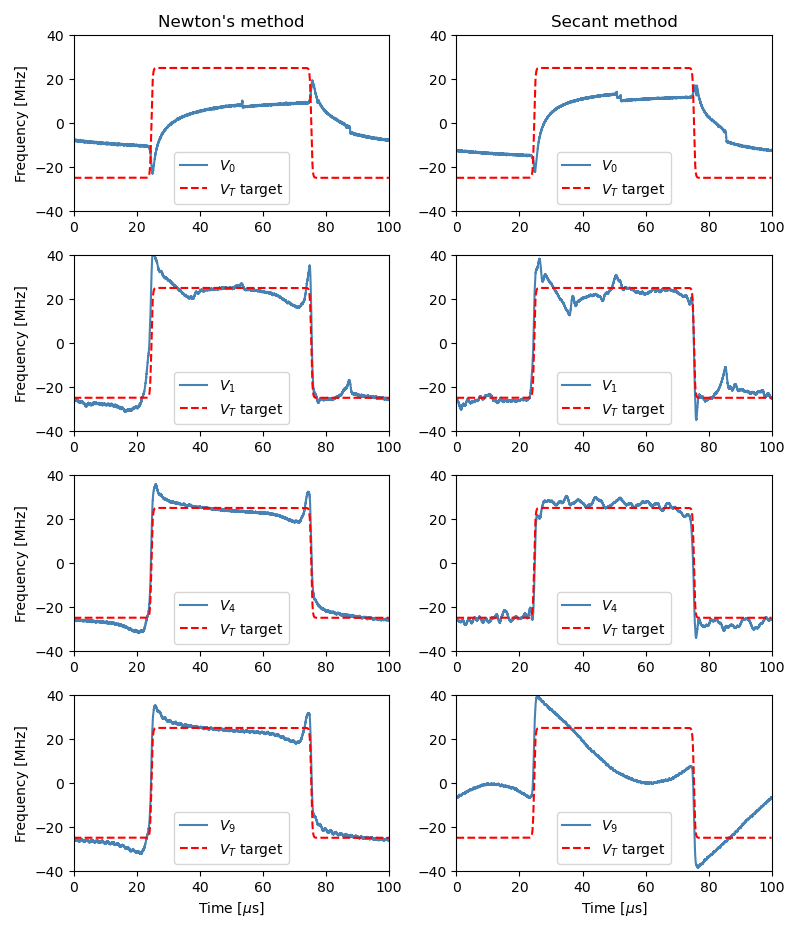} 
\caption{Poor converge of the algorithm after 10 iterations for the target FS pattern with a rise time $\tau=400$\,ns. I only represent the frequency response and not the input modulation signal $U_n$. The Newton's and secant methods are plotted on the left and right columns respectively.}
\label{fig:counterexample}
\end{figure}

For $\tau=400$ ns, the algorithm convergence is not convincing for the FS scheme regardless of the method used, as can be seen in Figure \ref{fig:counterexample}.

Both methods fail, but for different reasons that are worth discussing. Newton's method fails to compensate for transients produced by abrupt frequency changes. "Batman ears" persist on the profile despite iterations. More surprisingly, the secant method seems to approach the correct profile for $V_4$ but then deviates from it for $V_9$, revealing a large oscillation across the entire profile. Over-sensitivity to noise seems intrinsic to the target function for a rise time of $\tau=400$ ns, which is difficult to combat, but is expressed differently depending on the method used. I discuss possible options for compensating for these effects in the conclusion of the main article.

\section{Frequency discriminators}\label{appendix:discri}
I will take a moment here to describe the interferometer used to measure the laser frequency shift.
\subsection{Construction of a short-arm interferometer for frequency measurement}\label{appendix:interfero}

I choose to use a  Michelson fibred interferometer configuration whose arms are closed by two Faraday mirrors. As pointed out by Jiang {\it et al.} \cite{Jiang:10}, the output polarization is automatically aligned, leading to a maximally contrasted interference without any adjustment.

\begin{figure}[htbp]
\centering
\includegraphics[width=.85\columnwidth]{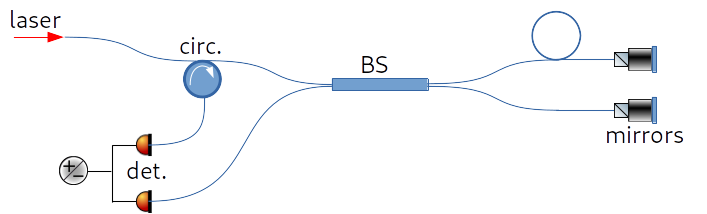} 
\caption{Short-arm fibred interferometer used as a frequency discriminator. Of Michelson type, one arm is slightly longer (about 5 cm) than the other, both are closed with  Faraday mirrors. Both outputs are analysed using a balanced detector ({\it det.}), one being collected using a fibred circulator ({\it circ.}). }
\label{fig:discriminator}
\end{figure}

I have decided to  use a short-arm interferometer (Fig. \ref{fig:discriminator}), in the sense that the free-spectral range (FSR) is greater than the modulation bandwidth. I take the time to justify this choice in \ref{appendix:short_long}. The difference in length between the two arms is only about 5 cm, adjusted by cutting the fibres before splicing. I then measure more precisely a FSR of 1.60\,GHz by scanning the laser and comparing it to a reference Fabry-Perot interferometer (Toptica FPI 100).

The frequency measurement is performed at the side of a fringe. It is then easy to eliminate intensity fluctuations by collecting and directly subtracting the two outputs of the interferometer on the balanced detector (Thorlabs PDB415A). This differential measurement eliminates the laser intensity fluctuations. When the modulation bandwidth is much smaller than the FSR, in other words, when we remain in the linear part of the fringe side, the voltage variations on the detector are proportional to the laser frequency.

For a short-arm interferometer (see \ref{appendix:short_long} for the discussion), it is necessary to actively maintain the laser on the fringe. I therefore use the differential signal, centred around zero, that provides the frequency measurement as an error signal to lock the laser using a PI servo controller (New Focus LB1005). I choose a slow response time (a few seconds) to avoid interfering with the modulation repetition rate $T=100\, \mu$s. This is a conservative choice, which decouples a slow lock-in and fast frequency modulation patterns under study, but this is not necessary. In fact, the locked laser as a whole constitutes a linear system, in other words with its own response function different from that of the free-running laser, which could be compensated for by the iterative algorithm. I have not carried out any tests in this regard, but it is conceivable.

\subsection{Short-arm or long-arm interferometer}\label{appendix:short_long}
I distinguish between two types of interferometers here: short-arm and long-arm, both of which can be used to measure frequency. I define short-arm when the FSR is greater than the bandwidth and vice versa for long-arm.

The advantage of a short-arm interferometer is that it directly gives the frequency when at the side of a fringe, as I explained in \ref{appendix:interfero}. No processing is necessary. The price to pay is the laser locking to keep it on the fringe. However, the complexity of the servo control should not be exaggerated, as it is slow in my case, especially since it is not specifically a matter of laser narrowing (noise reduction) but just a loose locking instead. A two-wave interferometer also has several locking points and a large capture range for the servo controller, unlike a cavity with large spans away from the resonance peaks without error signal to relock. That is the choice I made for those reasons.

Conversely, the long-arm interferometer does not require laser locking. Instead, during laser sweeps, the fringes are rapidly scanned to obtain a modulated homo- or heterodyne signal. The sweeping across a fringe then corresponds to the frequency scan of the interferometer FSR. To convert the modulated signal into a frequency measurement, processing is required, typically using a Hilbert transform. This is a disadvantage of the long-arm, especially since phase recovery processing leaves ambiguities depending on the phase wrapping or unwrapping.

\end{document}